\documentclass[prb,aps,amsmath,twocolumn,amssymb,floatfix,superscriptaddress]{revtex4}
\usepackage{amsmath}
\usepackage{latexsym}
\usepackage{amssymb}
\usepackage{graphics,epstopdf}
\usepackage[colorlinks=true, citecolor=blue, urlcolor=blue ]{hyperref}
\usepackage{epsf,graphics,graphicx}

\date{\today}

\begin{document}

\title{Localization to delocalization transition in a double stranded helical geometry: Effects of conformation, 
transverse electric field and dynamics}

\author{Suparna Sarkar}

\affiliation{Physics and Applied Mathematics Unit, Indian Statistical
Institute, 203 Barrackpore Trunk Road, Kolkata-700 108, India}

\author{Santanu K. Maiti}

\email{santanu.maiti@isical.ac.in}

\affiliation{Physics and Applied Mathematics Unit, Indian Statistical
Institute, 203 Barrackpore Trunk Road, Kolkata-700 108, India}

\begin{abstract}

Conformational effect on electronic localization is critically investigated for the first time considering a double-stranded helical 
geometry (DSHG) subjected to an electric field. In the presence of electric field the DSHG behaves like a correlated disordered 
system whose site potentials are modulated in a cosine form like the well known Aubry-Andr\'{e}-Harper (AAH) model. The potential 
distribution can be modulated further by changing the orientation of the incident field. A similar kind of cosine modulation is also
introduced in the inter-strand hopping integrals of the DSHG. Suitably adjusting the orientation of the electric field, we can achieve 
fully extended energy eigenstates or completely localized ones or a mixture of both. The effects of short-range and long-range hopping 
integrals along with the chirality on localization are thoroughly studied. Finally, we inspect the role of helical dynamics to make the 
model more realistic. The interplay between the helical geometry and electric field may open up several notable features of electronic 
localization and can be verified by using different chiral molecules.

\end{abstract}

\maketitle

\section{Introduction}

Helical structures always allow us to find new aspects due to their unique and diverse 
characteristics~\cite{hl1,hl2,hl3,hl4,hl5,hl6,hl7,hl8}. For instance, the observation of topological Hall effect~\cite{topohall} in 
helical spin geometries, the so-called helimagnetic materials~\cite{helimag}, detection of magnetic monopoles~\cite{magmp}, magnetic 
vortices~\cite{magv1,magv2}, and to name a few. Different kinds of simple and complex helimagnetic materials can be designed 
with the help of advanced nanofabrication techniques that reveal several non-trivial magnetic properties together with spin 
dynamics~\cite{dev1,dev2,dev3}. Along with these tailor made geometries we can find a biological world which is almost full with real 
helical structures. Several noteworthy features have already been put forward considering helical biological molecules~\cite{bio1,bio2,
bio3,bio4,bio5,bio6,bio7,bio8,bio9,bio10,bio11,maly}. The experimental demonstration of chiral induced spin selectivity~\cite{ss1} opens 
up the new possibilities of getting selective spin transmission through DNA and other helical molecules. Many other contemporary works 
have also been done along this line. At the same footing, finding of topological states in helical (organic) molecules is also very 
fascinating~\cite{topo1}, as it is directly related to other many physical systems.

So, undoubtedly helical systems bring significant impact and prove their robustness in different contexts. Although several aspects 
have been studied widely using helical structures, much attention has not been paid to explore electronic localization phenomenon which 
is one of the key issues to study transport properties. The classic problem of electronic localization is still alive since its prediction 
in 1958 by P. W. Anderson~\cite{andr1}. It is well known that for a 1D lattice with `uncorrelated' site potentials all energy eigenstates 
are exponentially localized~\cite{loc1,loc2,loc3,loc4}, irrespective of disorder strength $W$, which yields the critical disorder strength 
$W_c=0$. It therefore makes the system quite a trivial one, and here one cannot find any {\em mobility edge}~\cite{medas1} (ME), which 
separates a conducting zone from an insulating one. Setting a specific restriction on site potentials one can get a finite critical 
point ($W_c \ne 0$), though ME cannot be observed in strictly 1D case with nearest-neighbor hopping. A suitable example of it is the 
well known Aubry-Andre or Harper (AAH) model~\cite{abr} where site potentials are modulated in the quasiperiodic form like~\cite{zil} 
$W\cos(2\pi b n)$, where $n$ is the lattice site index and $b$ is an irrational number. In order to have ME, we need to couple at least 
two such 1D diagonal AAH chains, vertically as well as diagonally, to form a two-stranded AAH ladder. Because of finite diagonal hopping, 
two critical points are emerged which result a {\em mixed phase}~\cite{mphase} (MP) zone that is essential to get a ME in the energy 
spectrum~\cite{mphase,skm1}.

Though few proposals are available describing ME and other related phenomena in quasi-one-dimensional aperiodic 
lattices~\cite{medas1,mphase,skm1}, no effort has been made so far to unravel the localization properties considering (i) helicity of 
the geometry, (ii) interplay between short-range hopping (SRH) or long-range hopping (LRH) integrals and aperiodicity in site potentials 
and (iii) the helical dynamics. These are the key issues that we want to explore in our work. To substantiate these facts, we consider 
a double stranded helical geometry and investigate the localization phenomena of non-interacting electrons, in presence of a transverse 
electric field (see Fig.~\ref{model}). Due to this field, site energies of DSHG get modulated~\cite{maly,topo1,skm2,lhrh} which makes 
the system a `correlated' disordered one. On the other 
hand, to mimic this helical geometry with the real biological samples viz, double stranded DNA and other helical-like molecules, we 
include aperiodic modulation in inter-strand hopping integrals as these hopping integrals are no longer identical due to helicity. 
In each of these two modulations i.e., in site energies and inter-strand hopping integrals, a phase factor is associated referred as 
$\varphi_{\nu}$ and $\varphi_{\lambda}$ respectively, among which $\varphi_{\nu}$ can be regulated `externally' in a very simple way 
as it is directly involved with the orientation of external electric field~\cite{topo1,skm2,lhrh}. A correlation among them 
($\varphi_{\nu}$ and $\varphi_{\lambda}$) has an important role in electronic localization.

The present work deals with the following issues. (i) The critical roles of phase factors $\varphi_{\nu}$ and $\varphi_{\lambda}$ on 
electronic localization both for the right- and left-handed helical geometries, and how they are interrelated with the hopping integrals. 
For the right-handed SRH helix, all energy eigenstates 
are localized when the phase difference $\varphi_d$ ($=\varphi_{\nu}-\varphi_{\lambda}$) is zero or integer multiple of $\pi$ i.e., 
$\varphi_d=k\pi$ where $k=0$, $1$, $2$, etc. The situation becomes completely opposite when $\varphi_d$ becomes identical to 
$(2k+1)\pi/2$, and under this condition all the states are perfectly conducting. In the intermediate values of $\varphi_d$ 
mixed phase zone appears, and hence, mobility edges can be found. Thus, from one conducting state to another can be established by 
regulating the phase $\varphi_d$, which can be made quite easily by changing the orientation of the injected electric field. Due to 
longer range hopping, absolute localized states are no longer possible for the LRH helix even when $\varphi_d=k\pi$, rather we get 
mixed states, and, when the phase difference is equal to an odd integer multiple of $\pi/2$, all states become exactly conducting. 
For the left-handed DSHGs, analogous phenomena are observed under modified conditions of the phase factors that can be understood
from our detailed mathematical analysis. (ii) The effect of helical dynamics is critically investigated. The static picture does not 
always yield the complete scenario, as 
already put forward by several groups in different other contemporary works~\cite{hdn1,hdn2,hdn3,hdn4,hdn5,hdn6}, and thus, we need 
to focus beyond that to re-check whether any new physical phenomenon is achieved or not in our present context. 

Simulating the Hamiltonian of DSHG within a tight-binding (TB) framework we analyze the role of phase factors $\varphi_{\nu}$ and 
$\varphi_{\lambda}$ on localization-to-delocalization (LTD) transition. For some specific values of $\varphi_{\nu}$ and
$\varphi_{\lambda}$ the results are performed completely analytically, and an analytical prescription is always better to have a clear 
understanding. Later we provide numerical results for more general cases of $\varphi_{\nu}$ and $\varphi_{\lambda}$ which include 
average inverse participation ratio (AIPR)~\cite{iprth,iprdas}, average density of states (ADOS)~\cite{iprth} and electronic 
transmission probability~\cite{gfn1,gfn2,gfn3,gfn4}. Sandwiching DSHG among two contact electrodes, we determine transmission 
probabilities with the help of well known Green's function formalism~\cite{gfn1,gfn2,gfn3,gfn4}.

The rest of the paper is arranged as follows. In Sec. II, we describe the helical geometry and its TB Hamiltonian. Section III includes
the analytical treatment to characterize the localization properties under different conditions of $\varphi_{\nu}$ and $\varphi_{\lambda}$,
both for the right- and left-handed DSHGs. All the numerical results are presented in Sec. IV. Finally, in Sec. V we summarize our 
important findings.   

\section{Double stranded Helical geometry and TB Hamiltonian}

\begin{figure}[ht]
{\centering \resizebox*{4.3cm}{5.5cm}{\includegraphics{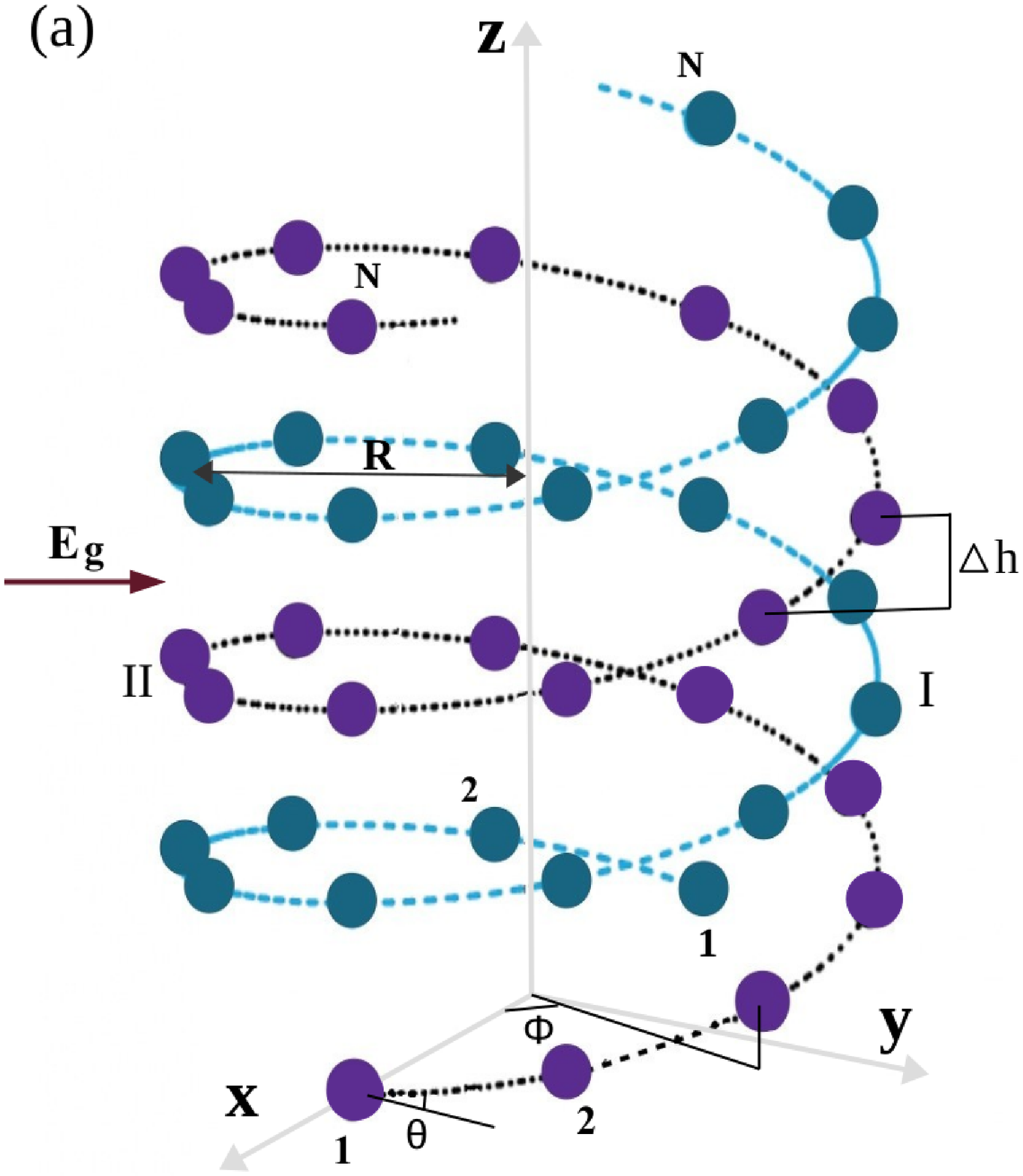}}
\raise 0.125cm \hbox{\kern 0.1cm \resizebox*{4cm}{5cm}
{\includegraphics{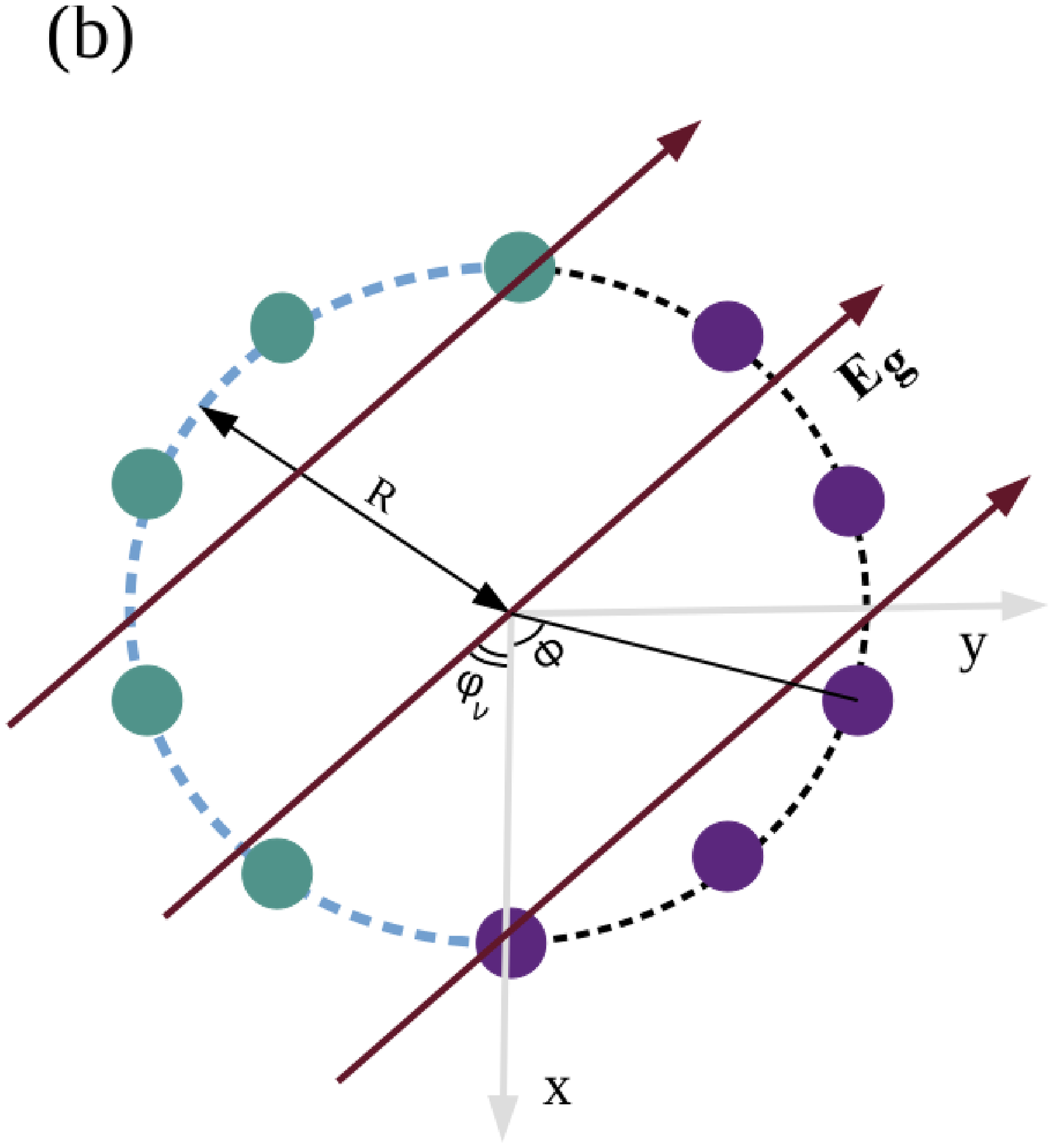}}}\par}
\caption{(Color online). (a) Sketch of a right-handed double stranded helical geometry in presence of an external electric
field of strength $E_g$. Here $R$ is the radius, $\Delta h$ is the stacking distance, $\theta$ is the helix angle and $\phi=n\Delta \phi$ 
($\Delta \phi$ being the twisting angle between the neighboring sites). (b) Projection of the bottom five base pairs and the external 
electric field in the $X$-$Y$ plane. $\varphi_{\nu}$ measures the angle between the positive $X$-axis and the incident electric field.
The direction of this field can be changed selectively.}
\label{model}
\end{figure}

We start by referring to Fig.~\ref{model}(a) where a schematic diagram of a right-handed double-stranded helical geometry is given. Each of 
the strands (I, II) contains $N$ lattice sites and they are arranged in a helical shape defined by the two fundamental factors, stacking 
distance $\Delta h$ and the twisting angle $\Delta \phi$ between the neighboring sites. For any $n$th site the angle $\phi$ 
(shown in Fig.~\ref{model}(a)) is defined as $\phi=n\Delta \phi$.
Depending on these parameters, we get the short-range or long-range hopping of electrons in the helical geometry. When $\Delta h$ is 
reasonably small i.e., the atoms are densely packed, longer range interactions are significant as electrons can hop into large enough 
distance. While for the other situation where $\Delta h$ is reasonably large, the atoms are separated far away which thus restricts the
motion of electrons into longer distances yielding a short-range hopping helix. These hopping integrals have direct correlation on
electronic localization, and in the present work we discuss all these issues one by one in a comprehensive way. In reality, two most
common examples those belong to the SRH and LRH groups are the DNA and protein molecules, respectively~\cite{bio9}. Thus, our analysis 
can directly be linked with these systems.

The Hamiltonian of the DSHG, shown in Fig.~\ref{model}(a), is simulated within a TB framework and it reads as
\begin{eqnarray}
H_{\mbox{\tiny }} & = & \sum_{j=\text {I,II}}\Big[\sum_{n=1}^{N} \epsilon_{j,n} c_{j,n}^{\dagger} c_{j,n} \nonumber \\
& + & \sum_{n=1}^{N-1}\sum_{i=1}^{N-n}t_{j,i}\Big(c_{j,n}^{\dagger} c_{j,n+i} + c_{j,n+i}^{\dagger}c_{j,n}\Big)\Big] \nonumber \\
& + & \sum_n\lambda_n\Big(c_{\tiny\mbox{I},n}^{\dagger} c_{\tiny\mbox{II},n} + c_{\tiny\mbox{II},n}^{\dagger} c_{\tiny\mbox{I},n}\Big)
\label{equ1}
\end{eqnarray}
where $j$ ($=\small\mbox{I},\small\mbox{II}$) represents the strand index, $n$ refers to the lattice sites in each strand, and
$c_{j,n}^{\dagger}$, $c_{j,n}$ are the usual fermionic operators. Different other factors of this TB Hamiltonian are described as follows. 

$t_{j,i}$ represents the intra-chain hopping integral between the sites $n$ and ($n+i$), and it becomes~\cite{bio10,skm2}
\begin{equation}
t_{j,i}=t_{j,1}e^{-\left(l_{j,i}-l_{j,1}\right)/l_c}
\label{hop}
\end{equation}
where $t_{j,1}$ is the nearest-neighbor hopping (NNH) strength, $l_c$ is the decay constant and $l_{j,i}$ measures the distance between 
the sites $n$ and ($n+i$). In terms of the radius $R$, stacking distance $\Delta h$ and twisting angle $\Delta \phi$, we can write 
$l_{j,i}$ as~\cite{bio10,skm2}
\begin{equation}
l_{j,i}=\sqrt{\left[2R \sin\left(\frac{i\Delta \phi}{2}\right)\right]^2 + \left(i\Delta h\right)^2}.
\label{lnt}
\end{equation}
$l_{j,i}$ maps to the nearest-neighbor distance when $i=1$. 

The parameter $\epsilon_{j,n}$ represents the site energy of an electron at site $n$ of the strand $j$. In the presence 
of external electric field $E_g$, perpendicular to the helix axis, the site energies in the two strands are~\cite{maly,topo1,lhrh}
\begin{equation}
\epsilon_{\tiny\mbox{I},n}=-\epsilon_{\tiny\mbox{II},n}=eV_g \cos \left(n\Delta \phi-\varphi_{\nu}\right)
\label{steq}
\end{equation}
where $e$ is the electronic charge, and $V_g$ is the gate voltage associated with the electric field by the relation $2V_g=2E_g R$.
From the projected view of the base pairs (see Fig.~\ref{model}(b)), the sign reversal of site energies in the two strands of the
right-handed DSHG (a realistic example of such a system can be the B form of DNA molecule~\cite{bdna}) can be understood, as the 
strands run in opposite directions. For more details see~\cite{maly}.
The phase factor $\varphi_{\nu}$ denotes the angle between the positive $X$-axis and the incident electric field (see Fig.~\ref{model}(b)), 
and it can be tuned by altering the field direction. The above form of site energies looks similar to the well known diagonal AAH 
model~\cite{zil} by imagining that $\Delta \phi$ is equivalent to the factor $2\pi b$ ($b$ being an irrational number). Thus, selectively 
choosing $\Delta \phi$ we can get a deterministic DSHG whose site energies are correlated in the form of AAH type, when the DSHG is 
subjected to the electric field $E_g$.

The rest other factor $\lambda_n$ of Eq.~\ref{equ1} describes the inter-strand coupling. We introduce a similar kind of cosine 
modulation in $\lambda_n$ and it is described as
\begin{equation}
\lambda_n=W \cos \left(n \Delta \phi^{\prime}-\varphi_{\lambda}\right)
\label{interhop}
\end{equation}
where $W$ is the cosine modulation strength and $\varphi_{\lambda}$ is the phase factor. $\Delta \phi^{\prime}$ is analogous to 
$\Delta \phi$.

Thus, two phase factors ($\varphi_{\nu}$ and $\varphi_{\lambda}$) are associated with DSHG and we concentrate on the interplay between 
them on electronic localization. Among these two phases, $\varphi_{\nu}$ can be tuned in a simple way by altering the field direction, 
as already stated. Now, keeping the phase factor $\varphi_{\lambda}$ constant, associated with inter-strand hopping integrals, we 
can get different conducting behaviors like fully conducting, or fully localized or the mixture of both by adjusting the other phase 
factor $\varphi_{\nu}$ satisfying some typical relations among $\varphi_{\nu}$ and $\varphi_{\lambda}$. Most importantly all these 
conducting features are found in a single system. The specific signatures of different conducting phases can be understood from our 
subsequent discussion.

\section{Characterization of localization properties: Analytical treatment}

Following an analytical treatment, in this section, we investigate the precise roles of $\varphi_{\nu}$ and $\varphi_{\lambda}$, and the 
interplay between them on electronic localization. Analytic prescriptions are always more useful to understand the basic physics behind
any phenomenon with better clarity. Both the right- and left-handed helical geometries are taken into account, and we characterize their 
features one by one as follows. In our analysis, we use bold face to denote a matrix.

\subsection{Right-handed double-stranded helical geometry}

Let us begin with the Schr\"{o}dinger equation 
\begin{equation}
H|\Psi\rangle=E|\Psi\rangle
\label{schequ1}
\end{equation}
 where $|\Psi\rangle=\sum_{j,n} \psi_{j,n} |j,n\rangle$.
$|j,n\rangle$'s are the Wannier functions and $\psi_{j,n}$'s are the co-efficients. Using the Schr\"{o}dinger equation we can write 
difference relation for any site $n$ of strand I as 
\begin{eqnarray}
\left(E-\epsilon_{\tiny\mbox{I},n}\right) \psi_{\tiny\mbox{I},n} = \sum_{i} t_{\tiny\mbox{I},i} \left(\psi_{\tiny\mbox{I},n+i}
+\psi_{\tiny\mbox{I},n-i}\right) + \lambda_n \psi_{\tiny\mbox{II},n}.
\label{equ2}
\end{eqnarray}
Similarly for the site $n$ of strand II, the difference equation becomes
\begin{eqnarray}
\left(E-\epsilon_{\tiny\mbox{II},n}\right)\psi_{\tiny\mbox{II},n} = \sum_{i} t_{\tiny\mbox{II},i} \left(\psi_{\tiny\mbox{II},n+i}
+ \psi_{\tiny\mbox{II},n-i}\right) + \lambda_n \psi_{\tiny\mbox{I},n}.
\label{equ3}
\end{eqnarray}
Combining Eq.~\ref{equ2} and Eq.~\ref{equ3} we get the matrix equation
\begin{eqnarray}
\left[\begin{pmatrix}
E & 0 \\
0 & E
\end{pmatrix}
-\begin{pmatrix}
\epsilon_{\tiny\mbox{I},n} & \lambda_n\\
\lambda_n & \epsilon_{\tiny\mbox{II},n}
\end{pmatrix}\right]
\begin{pmatrix}
\psi_{\tiny\mbox{I},n} \\
\psi_{\tiny\mbox{II},n}
\end{pmatrix} & = &
\nonumber \\
\nonumber\\
\sum_{i} \begin{pmatrix}
t_{\tiny\mbox{I},i} & 0\\
0 & t_{\tiny\mbox{II},i}
\end{pmatrix}
\left[\begin{pmatrix}
\psi_{\tiny\mbox{I},n+i} \\
\psi_{\tiny\mbox{II},n+i}
\end{pmatrix}
+\begin{pmatrix}
\psi_{\tiny\mbox{I},n-i} \\
\psi_{\tiny\mbox{II},n-i}
\end{pmatrix}\right]
\label{equ4}
\end{eqnarray}
which can be re-written in a compact way as
\begin{eqnarray}
\left(\mbox{\boldmath{$E$}}-\mbox{\boldmath{$M_n$}}\right)\mbox{\boldmath$\psi_{n}$} = \sum_{i} \mbox{\boldmath{$t_i$}}
\left(\mbox{\boldmath$\psi_{n+i} + \psi_{n-i}$}\right)
\label{equ5}
\end{eqnarray}
where the matrix forms of different matrices can easily be followed. The non-zero off-diagonal terms in \mbox{\boldmath{$M_n$}} restricts
the decoupling of Eq.~\ref{equ5} into two separate 1D chains. We can construct a matrix \mbox{\boldmath{$S_n$}} such that it 
diagonalizes \mbox{\boldmath{$M_n$}} i.e, \mbox{\boldmath{$S_n^{-1}M_nS_n=M_n^d$}}, where \mbox{\boldmath{$M_n^d$}} is the diagonal one.
Doing some mathematical steps we find
\begin{eqnarray}
\mbox{\boldmath{$S_n$}}=\begin{pmatrix}
\frac{\alpha}{\sqrt{1+\alpha^2}} & \frac{\beta}{\sqrt{1+\beta^2}}\\
\frac{1}{\sqrt{1+\alpha^2}} & \frac{1}{\sqrt{1+\beta^2}}
\end{pmatrix}
\label{equ6} 
\end{eqnarray}
where,
\begin{subequations}
\begin{align}
\alpha &=\frac{(\epsilon_{\tiny\mbox{I},n}-\epsilon_{\tiny\mbox{II},n}) - \sqrt{(\epsilon_{\tiny\mbox{I},n} -\epsilon_{\tiny\mbox{II},n})^2
+4 \lambda_n^2}}{2\lambda_n} \\
\beta &=\frac{(\epsilon_{\tiny\mbox{I},n}-\epsilon_{\tiny\mbox{II},n}) + \sqrt{(\epsilon_{\tiny\mbox{I},n}-\epsilon_{\tiny\mbox{II},n})^2
+4 \lambda_n^2}}{2\lambda_n}.
\label{equ7}
\end{align}
\end{subequations}
Here it is important to note that the matrix \mbox{\boldmath{$S_n$}} is no longer constant i.e., not independent of the site index $n$, 
as the matrix elements are explicit functions of $\epsilon_{\tiny\mbox{I},n}$ and $\epsilon_{\tiny\mbox{II},n}$.
Plugging \mbox{\boldmath{$S_n$}} and \mbox{\boldmath{$S_n^{-1}$}} appropriately in Eq.~\ref{equ5}, we get the matrix equation
\begin{equation}
\left(\mbox{\boldmath{$E-M_n^d$}}\right)\mbox{\boldmath $\chi_{n,n}$}=\sum_{i} \mbox{\boldmath{$t_i$}} 
\left(\mbox{\boldmath$\chi_{n,n+i} + \chi_{n,n-i}$}\right)
\label{equ9}
\end{equation}
where,
\begin{equation}
\mbox{\boldmath$\chi_{n,n}$} = \mbox{\boldmath{$S_n^{-1} \psi_n$}} = 
\begin{pmatrix}
-\sqrt{\frac{1+\alpha^2}{4+\gamma^2}} \psi_{\tiny\mbox{I},n} + \frac{1}{\sqrt{2+\alpha\gamma}} \psi_{\tiny\mbox{II},n}\\
\sqrt{\frac{1+\beta^2}{4+\gamma^2}} \psi_{\tiny\mbox{I},n} + \frac{1}{\sqrt{2+\beta\gamma}}\psi_{\tiny\mbox{II},n}
\end{pmatrix}
\label{equ10}
\end{equation}
with 
\begin{equation}
\gamma=\frac{\epsilon_{\tiny\mbox{I},n} -\epsilon_{\tiny\mbox{II},n}}{\lambda_n}.
\end{equation}
The elements of \mbox{\boldmath$\chi_{n,n}$} also depend on $\epsilon_{\tiny\mbox{I},n}$ and $\epsilon_{\tiny\mbox{II},n}$.
The diagonal elements of \mbox{\boldmath{$M_n^d$}} are
\begin{subequations}
\begin{align}
\mbox{\boldmath$M_n^d$}_{11} &{\small =\frac{1}{2}\left[\epsilon_{\tiny\mbox{I},n}+\epsilon_{\tiny\mbox{II},n}
-\sqrt{(\epsilon_{\tiny\mbox{I},n}-\epsilon_{\tiny\mbox{II},n})^2 + 4\lambda_n^2}\right]} \label{equ11} \\
\mbox{\boldmath$M_n^d$}_{22} &{\small =\frac{1}{2}\left[\epsilon_{\tiny\mbox{I},n}+\epsilon_{\tiny\mbox{II},n}
+\sqrt{(\epsilon_{\tiny\mbox{I},n}-\epsilon_{\tiny\mbox{II},n})^2+4\lambda_n^2}\right].}
\label{equ111}
\end{align}
\end{subequations}
We have now all the required basic expressions in our hand, and thus, different situations of DSHG can be achieved depending on the 
input conditions, and they are as follows.

\vskip 0.2cm
\noindent
Case I: \underline{\em DSHG without external electric field}
\vskip 0.2cm

In the absence of any electric field $\epsilon_{\tiny\mbox{I},n}$ becomes identical with $\epsilon_{\tiny\mbox{II},n}$. Under this 
condition $\mbox{\boldmath{$S_n$}}$ and $\boldsymbol{\chi}_{n,n}$ become independent of $\alpha$, $\beta$ and $\gamma$ terms, and they are 
simplified as
\begin{equation}
\mbox{\boldmath{$S_n$}}=\mbox{\boldmath{$S$}}=\frac{1}{\sqrt{2}}
\begin{pmatrix}
-1 & 1\\
1 & 1
\end{pmatrix}
\label{ffeq17}
\end{equation}
and
\begin{equation}
\mbox{\boldmath$\chi_n$}=\mbox{\boldmath{$S^{-1}\psi_n$}}=\frac{1}{\sqrt{2}}
\begin{pmatrix}
-\psi_{\tiny\mbox{I},n}+\psi_{\tiny\mbox{II},n}\\
\psi_{\tiny\mbox{I},n}+\psi_{\tiny\mbox{II},n}
\end{pmatrix} \equiv
\begin{pmatrix}
\chi_{\tiny\mbox{I},n} \\
\chi_{\tiny\mbox{II},n}
\end{pmatrix}.
\label{ffeq18}
\end{equation}
For this typical case, the DSHG can be decoupled into two 1D chains those satisfy the following difference equations.
\begin{subequations}
\begin{align}
\left(E+\lambda_n\right)\chi_{\tiny\mbox{I},n} & =\sum_{i} t_{\tiny\mbox{I},i} \left(\chi_{\tiny\mbox{I},n+i}
+ \chi_{\tiny\mbox{I},n-i}\right) \label{equ13} \\
\left(E-\lambda_n\right)\chi_{\tiny\mbox{II},n} & =\sum_{i} t_{\tiny\mbox{II},i} \left(\chi_{\tiny\mbox{II},n+i}
+\chi_{\tiny\mbox{II},n-i}\right).
\label{equ14}
\end{align}
\end{subequations}
These two decoupled chains (Eqs.~\ref{equ13} and \ref{equ14}) are analogous to the well known AAH chains with higher order hopping
integrals where AAH potentials are incorporated in site energies. For these systems, localization phenomena have already been studied
at some level. Moreover, in the absence of electric field one cannot examine the interplay between the phases $\varphi_{\nu}$ and 
$\varphi_{\lambda}$, as $\varphi_{\nu}$ does not exist in the absence of electric field. Therefore, here we do not focus on this field 
free case, rather we concentrate on the results of non-zero electric field as given below.

\vskip 0.2cm
\noindent
Case II: \underline{\em DSHG with external electric field}
\vskip 0.2cm

In the presence of electric field $\epsilon_{\tiny\mbox{I},n}$ and $\epsilon_{\tiny\mbox{II},n}$ are no longer identical, rather they
are connected as $\epsilon_{\tiny\mbox{I},n}=-\epsilon_{\tiny\mbox{II},n}$ (see Eq.~\ref{steq}). Therefore, \mbox{\boldmath$S_n$} and
\mbox{\boldmath$\chi_{n,n}$} cannot get their simplified forms as described for the field free case (Eqs.~\ref{ffeq17} and \ref{ffeq18}),
while they depend on $\alpha$, $\beta$ and $\gamma$ as illustrated in Eqs.~\ref{equ6} and \ref{equ10}. Under this case, it is not 
possible to decouple the DSHG into two decoupled chains. The primary condition to have decoupled chains from a 
two-stranded ladder is that the site energies of each pair those are connected vertically i.e., $\epsilon_{\tiny\mbox{I},n}$ and 
$\epsilon_{\tiny\mbox{II},n}$ should be identical with each other, though for different pairs (viz, $n\ne m$) they can be different.
This idea was originally put forward in Ref.~\cite{skm1} with detailed mathematical description.

Now, we examine the interplay between $\varphi_{\nu}$ and $\varphi_{\lambda}$ on electronic localization. 
For $\epsilon_{\tiny\mbox{I},n}=-\epsilon_{\tiny\mbox{II},n}$, Eqs.~\ref{equ11} and \ref{equ111} get simplified as 
\begin{subequations}
\begin{align}
\mbox{\boldmath$M_n^d$}_{11} & = -\sqrt{\epsilon_{\tiny\mbox{I},n}^2 + \lambda_n^2} \label{equ20a} \\
\mbox{\boldmath$M_n^d$}_{22} & = \sqrt{\epsilon_{\tiny\mbox{I},n}^2 + \lambda_n^2}.
\label{equ20b}
\end{align}
\end{subequations}
Adjusting $V_g$ appropriately we can fix $eV_g=W$ (identical modulation strength in both sectors), and we also set 
$\Delta \phi^{\prime}=\Delta \phi$. For these choices we get 
\begin{subequations}
\begin{align}
\mbox{\boldmath$M_n^d$}_{11} & = -W \sqrt{\cos^2(n\Delta\phi-\varphi_\nu)+\cos^2(n\Delta\phi-\varphi_\lambda)} \label{equ21a} \\
\mbox{\boldmath$M_n^d$}_{22} & = W \sqrt{\cos^2(n\Delta\phi-\varphi_\nu)+\cos^2(n\Delta\phi-\varphi_\lambda)}.\label{equ21b}
\end{align}
\end{subequations}
Now, if the phase difference $\varphi_d$ ($=\varphi_{\nu}-\varphi_{\lambda}$) is an `odd integer' multiple of $\pi/2$ i.e.,
$\varphi_d=(2k+1)\pi/2$, where $k$ being an integer, then the above two matrix elements boil down to
\begin{equation}
\mbox{\boldmath$M_n^d$}_{11} = -W ~~~ \mbox{and} ~~~ \mbox{\boldmath$M_n^d$}_{22} = W.
\label{equn22}
\end{equation} 
Thus, the diagonal elements of \mbox{\boldmath$M_n^d$} are independent of $n$, or, more precisely we can say that they are independent 
of cosine modulations. Therefore, for a DSHG subjected to an external electric field, we can eventually reach to a situation by adjusting 
the phase $\varphi_{\nu}$, where the effect of cosine modulations gets completely eliminated. This leads to a fully perfect DSHG, where 
all the eigenstates will be extended. For all other phase factors, DSHG behaves as a correlated disordered one satisfying the 
cosine modulations as described above, and depending on the hopping integrals (SRH or LRH ones) we can get fully localized or mixture of 
both extended and localized states. These issues are discussed in the subsequent parts. 

\subsection{Left-handed double-stranded helical geometry}

Now, we focus on the left-handed double-stranded helical geometry where both the strands are left handed (schematic 
diagram is not shown here as it can easily be understood). A suitable realistic example of such a system is Z form of DNA~\cite{zdna}.
The change of helicity of DSHG from the right-handed to left-handed one is described in the TB Hamiltonian by 
$\Delta \phi \rightarrow -\Delta \phi$~\cite{lhrh}. 
The sign reversal in $\Delta \phi^{\prime}$, associated with inter-strand hopping, does not take place with the change in helicity. 
Thus, we have the forms of site energies and inter-strand hopping integrals as
\begin{eqnarray}
\epsilon_{\tiny\mbox{I},n} = - \epsilon_{\tiny\mbox{II},n} = W \cos \left(n\Delta \phi+\varphi_\nu \right)
\label{equ23}
\end{eqnarray} 
and 
\begin{eqnarray}
\lambda_n=W \cos \left(n\Delta \phi^{\prime} - \varphi_\lambda\right).
\label{equ24}
\end{eqnarray}
Substituting these factors in Eqs.~\ref{equ11} and \ref{equ111} we get (assuming $\Delta\phi^{\prime}=\Delta \phi$, like above)
\begin{subequations}
\begin{align}
\mbox{\boldmath$M_n^d$}_{11} & = -W \sqrt{\cos^2(n\Delta\phi+\varphi_\nu)+\cos^2(n\Delta\phi-\varphi_\lambda)} \label{equ25a} \\
\mbox{\boldmath$M_n^d$}_{22} & = W \sqrt{\cos^2(n\Delta\phi+\varphi_\nu)+\cos^2(n\Delta\phi-\varphi_\lambda)}.\label{equ25b}
\end{align}
\end{subequations}
Due to a sign change in arguments in the two cosine modulations (see Eqs.~\ref{equ23} and \ref{equ24}), we need to satisfy the following
two conditions to have the absolute conducting phase, as clearly noticed from the expressions given in Eqs.~\ref{equ25a} and 
\ref{equ25b}. The conditions are
\begin{subequations}
\begin{align}
\varphi_d=\varphi_{\nu}-\varphi_{\lambda} & = (2k+1) \frac{\pi}{2} \label{equ26a} \\
\varphi_{\lambda} = k \frac{\pi}{2}~~&\mbox{or}~~ \varphi_{\nu} = k \frac{\pi}{2}. \label{equ26b}
\end{align}
\end{subequations}
Along with Eq.~\ref{equ26a}, Eq.~\ref{equ26b} has to be imposed unlike the right-handed DSHG. Under these conditions of $\varphi_{\nu}$ 
and $\varphi_{\lambda}$, the matrix elements $\mbox{\boldmath$M_n^d$}_{11}$ and 
$\mbox{\boldmath$M_n^d$}_{22}$ become $n$ independent like right-handed DSHG and they are $-W$ and $W$ respectively. This leads to the 
absolute conducting behavior for the left-handed DSHG. Deviating from it will yield the other conducting phases (see forthcoming analysis).

We end our analytical description by stating that it is done for arbitrary range of electron hopping in the DSHG i.e.,
be it short-range or long-range one.

\section{Numerical Results}

From the above analytical description it is clearly understood that under certain conditions a DSHG, subjected to an external electric 
field, can exhibit a perfectly conducting phase where all the energy eigenstates are extended. But for arbitrary choices of 
$\varphi_{\nu}$, $\varphi_{\lambda}$ and other physical parameters describing the geometry, analytical approach is almost impossible, 
and hence we provide numerical results for the completeness of our analysis. The numerical results include average density of states 
(ADOS) together with two-terminal transmission probability ($T$)~\cite{gfn1,gfn2,gfn3,gfn4} under different input conditions.
Along with these, average inverse participation ratio (AIPR)~\cite{iprth,iprdas} is also analyzed, which is another important tool to 
examine the nature of conducting behavior of any system. 

To find transmission probability, a finite size DSHG is clamped between two electronic reservoirs those are usually called as source (S)
and drain (D). Following the well known Green's function prescription, transmission probability is obtained from the 
relation~\cite{gfn1,gfn2,gfn3} 
\begin{equation}
T=\mbox{Tr}\left[\Gamma_S G^r \Gamma_D G^a \right]
\label{trp}
\end{equation}
where $G^r$ and $G^a$ are the retarded and advanced Green's functions, respectively, and $\Gamma_S$ and $\Gamma_D$ are the coupling 
factors. In terms of the self-energies $\Sigma_S$ and $\Sigma_D$ due to S and D, the Green's functions are defined as~\cite{gfn2,gfn3}
\begin{equation}
G^r=(G^a)^{\dagger}=\frac{1}{E-H-\Sigma_S-\Sigma_D}
\label{grf}
\end{equation}
and the coupling terms are obtained from the expression $\Gamma_{S(D)}=-2\mbox{Im}[\Sigma_{S(D)}]$. Using the Green's function,
ADOS is also determined through the relation
\begin{equation}
\rho(E)=-\frac{1}{2N\pi} \mbox{Im}[\mbox{Tr}(G^r)]
\label{dos}
\end{equation}  
where $2N$ represents the total number of lattice sites in DSHG. In our work, we simulate the source and drain electrodes in 
the form of a 1D chain, parameterized by on-site energy $\epsilon_0$ and NNH integrals $t_0$. These electrodes are coupled to
the DSHG via the coupling strengths $t_S$ and $t_D$.

Inverse participation ratios (IPRs) are evaluated from the normalized eigenvectors. To have the full picture, we find IPRs of 
the DSHG considering the effects of the contact electrodes i.e., for the effective Hamiltonian 
$H_{\mbox{\tiny eff}}=H+\Sigma_S+\Sigma_D$. If $|\Psi^p\rangle$ ($=\sum_{j,n}\psi_{j,n}^p|j,n\rangle$) is a normalized eigenstate
($p$ being the state index), then IPR for this state is described by~\cite{iprth,iprdas}
\begin{equation}
IPR_p=\sum_{j,n} |\psi_{j,n}^p|^4.
\label{ipr30}
\end{equation}
Calculating IPRs for all the $2N$ states, we compute AIPR of the system which is~\cite{iprdas} 
\begin{equation}
AIPR=\frac{1}{2N} \sum_p IPR_p.
\label{aipr31}
\end{equation}
For a fully perfect system where all the states are extended $AIPR \rightarrow 0$, while $AIPR \rightarrow 1$ when all the
\begin{table}[ht]
\caption{Structural parameters of LRH and SRH helical geometries.}
\vskip 0.15cm
\begin{tabular}{|c|c|c|c|c|c|}
\hline \hline
Hopping & $R$ ($\mbox\AA$) & $\Delta h$ ($\mbox\AA$) & $\Delta\phi$ & $l_c$ ($\mbox\AA$) & $b$ \\ 
\hline 
SRH & $8$ & $4.3$ & $\frac{\pi(\sqrt{5}-1)}{4}$ & $0.8$ & $\frac{\sqrt{5}-1}{8}$ \\ 
\hline 
LRH & $2.4$ & $1.6$ & $\frac{\pi(\sqrt{5}-1)}{2}$ & $1.2$ & $\frac{\sqrt{5}-1}{4}$ \\ 
\hline \hline 
\end{tabular} 
\label{tab1}
\end{table}
states get absolutely localized~\cite{iprdas}. For finite size system, `absolute' localization i.e., where $AIPR \rightarrow 1$, is
quite hard to achieve, but it can reach to a moderate value ($>0.5$ or even more). In both these two cases there is no possibility to 
have a mobility edge (ME). For smaller values of AIPR (but not very close to zero), we get a mixture of both conducting and localized 
states~\cite{iprdas}. Under this situation, mobility edge phenomenon can be observed. In our numerical analysis, we confirm the existence
of the ME in two ways. One by inspecting transmission probability together with the ADOS spectrum, and in the other way by examining
the behavior of AIPR.

Before presenting the numerical results, let us briefly mention the common parameters values. To implement SRH and LRH geometries, we 
choose the physical parameters analogous to the real biological systems~\cite{dnapr}, DNA and $\alpha$-helical protein, those are ideal
and established examples of SRH and LRH systems, as put forward by different groups. The structural parameters of the two helical 
geometries (LRH and SRH) are given in Table~\ref{tab1}. From the relation $\Delta \phi=2\pi b$, we get the above values of $b$ 
\begin{figure}[ht]
{\centering \resizebox*{8.4cm}{8cm}{\includegraphics{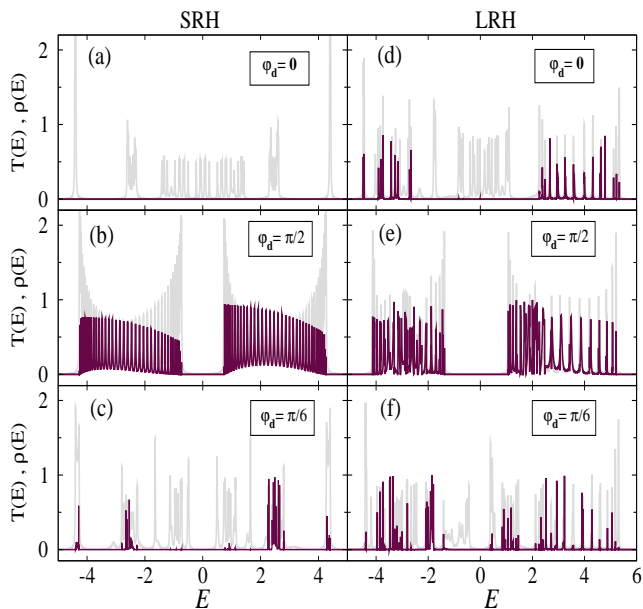}}\par}
\caption{(Color online). Right-handed DSHG: Transmission probabilities (maroon color) and density of states (light gray color) of SRH 
and LRH DSHGs at some typical values of $\varphi_d$ when $V_g$ is fixed at $2.5\,$V. Here we choose $N=40$.}
\label{tran}
\end{figure}
(given in the right column of Table~\ref{tab1}) for the SRH and LRH cases, those are incommensurate. The other common physical
parameters are: $\epsilon_0=0$, $t_0=3$, $t_S=t_D=t_1=1$. All the energies are measured in unit of electron-volt (eV). 
Both for the right-handed and left-handed helical geometries, the numerical results are presented one by one.

\subsection{Right-handed DSHG}

In Fig.~\ref{tran} we show transmission probabilities of the SRH and LRH DSHGs at three distinct values of $\varphi_d$. In each 
of these spectra, ADOS is superimposed to get a clear picture of the allowed energy channels. We take a reasonably large $V_g$, such 
that the cosine modulation strength $W$ ($=eV_g$) becomes much higher compared to $t_1$. Several interesting patterns are emerged.
For the short-range helix geometry, transmission probability vanishes completely for the entire energy window when the phase difference 
is zero (Fig.~\ref{tran}(a)). It indicates that all the energy eigenstates are localized and thus the system goes to the insulating state. 
This is due to the well known effect of disorder in different site energies of the two strands and modulated inter-strand hopping integrals.
\begin{figure}[ht]
{\centering \resizebox*{8cm}{10cm}{\includegraphics{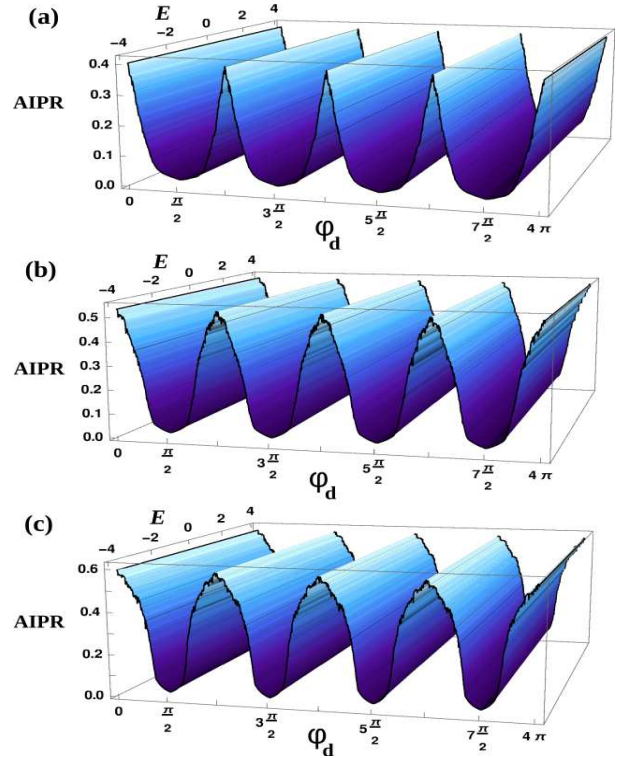}}\par}
\caption{(Color online). Right-handed DSHG: 3D plot of AIPR as functions of phase difference $\varphi_d$ and electronic energy $E$ for
the identical set of parameter values as taken in Fig.~\ref{tran}, where (a), (b) and (c) correspond to $V_g=3$, $5$ and $7\,$V,
respectively. The results are computed for the SRH DSHG.}
\label{iprplot}
\end{figure}
The situation becomes completely opposite when the phase difference $\varphi_d=\pi/2$. Absolute conducting state is obtained 
(Fig.~\ref{tran}(b)) as for all the energy channels we get finite transmission. At this typical value of $\varphi_d$, the cosine 
modulations get washed out completely, that can be clearly visible from our mathematical analysis, and the system behaves like a perfect 
double-stranded helical conductor. For the intermediate value of $\varphi_d$ i.e., $0 < \varphi_d < \pi/2$ (here we set $\varphi_d=\pi/6$, 
as a typical example) interestingly we find that there are some energy windows for which finite transmission takes place 
(Fig.~\ref{tran}(c)), while for the rest of the energy channels transmission probability vanishes. This is a clear indication of the 
existence of both extended and localized energy eigenstates in the allowed energy zone. A careful inspection reveals that few discrete 
energies are available whose one side exhibits finite transmission, while vanishing transmission is obtained for the other side 
(Fig.~\ref{tran}(c)). Thus, a seperation between conducting and insulating states exists accross these typical energies, and these 
energies are referred as mobility edges.   

In the case of long-range hopping the situation is little bit different. Unlike the SRH DSHG, finite transmission is available for 
some energy levels even when the phase difference $\varphi_d=0$ (Fig.~\ref{tran}(d)). This is because of the longer range hopping of 
electrons which prevent electrons to get localized even when the impurity strength is too high. Of course the transmission probability 
decreases with increasing the modulation strength (here it is the strength of electric field $V_g$), but absolute localization for finite 
size system with longer range hopping integrals is hard to achieve. As long as the phase difference $\varphi_d$ is set equal to $\pi/2$, 
all the energy levels become fully conducting and electrons can pass through the DSHG from source to the drain end
(Fig.~\ref{tran}(e)). For $\varphi_d=\pi/6$, more conducting channels are obatined (Fig.~\ref{tran}(f)) comapred to $\varphi_d=0$ case.

Here it is relevant to note that even when the modulation strength $W$ is too high, the absolute conducting state is always obtained at 
$\varphi_d=\pi/2$ since for this situation cosine modulations no longer persist. This feature remains
\begin{figure}[ht]
{\centering \resizebox*{8.4cm}{8cm}{\includegraphics{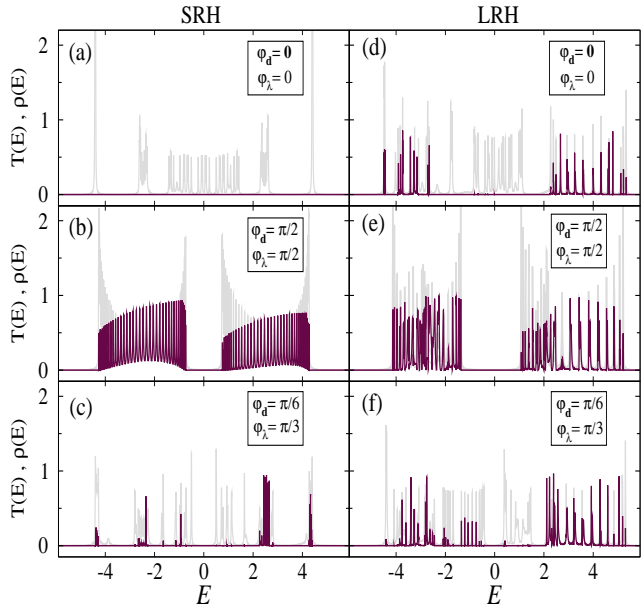}}\par}
\caption{(Color online). Left-handed DSHG: Transmission probabilities (maroon color) and density of states (light gray color) of 
SRH and LRH DSHGs at some typical values of $\varphi_d$ and $\varphi_{\lambda}$. The system size and gate voltage remain same as 
considered in Fig.~\ref{tran}.}
\label{lhtran}
\end{figure}
unchanged when $\varphi_d$ becomes identical to any odd integer multiple of $\pi/2$, which we verify through our numerical calculations. 
The numerical results fully corroborate our analyical findings. Thus, for the SRH DSHG, a complete localization-to-delocalization
(LTD) transition can be established by regulating the phase factors, or more suitably by tuning the orientation of external electric 
field when the other phase $\varphi_{\lambda}$ remains fixed. Moreover, mixture of localized and extended states is also available by 
controlling the phases associated with the cosine modulations.

To visualize more precisely the phenomenon of LTD transition at some typical values of $\varphi_d$, and, the appearance of mixed phase 
at intermediate $\varphi_d$'s, in Fig.~\ref{iprplot} we show the variation of AIPR by tuning $\varphi_d$ in a wide range. AIPR gives a 
very good estimate for determining the conducting properties of different energy states as it is directly involved with the participation 
of electrons at distinct lattice sites. The results are presented for the SRH DSHG, since localized phase is clearly noticed due to 
short-range hopping of electrons. A regulatory oscillating pattern is exhibited with $\varphi_d$, providing a minimum 
(AIPR $\rightarrow 0$) at $\varphi_d=(2k+1)\pi/2$, yielding the conducting phase, while AIPR becomes maximum at $\varphi_d=k\pi$ which 
suggests the localized phase. AIPR $\rightarrow 1$ is obtained only in the asymptotic limit where $N\rightarrow \infty$. For the 
intermediate values of $\varphi_d$, the signature of mixed energy eigenstates is reflected.
All these issues are fully consistent with our analytical findings.

\subsection{Left-handed DSHG}

For the sake of completeness and to understand the left-right correspondence,
\begin{figure}[ht]
{\centering \resizebox*{8.4cm}{8cm}{\includegraphics{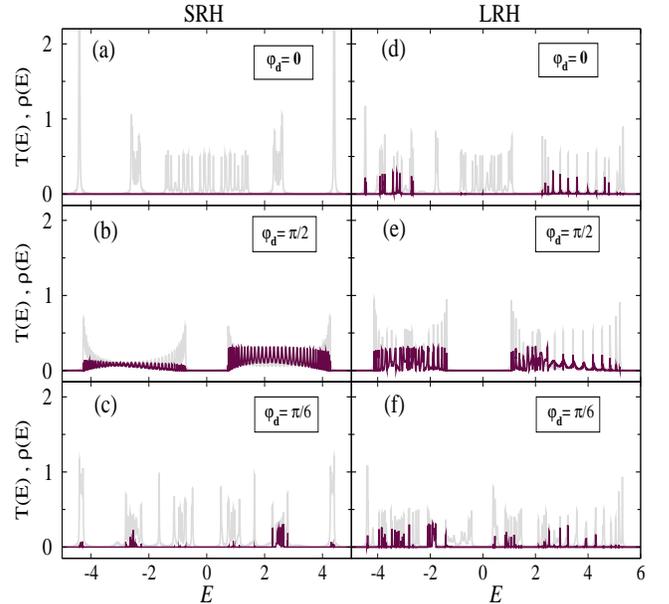}}\par}
\caption{(Color online). Effect of helical dynamics in right-handed DSHG is considered. The transmission probabilities and ADOS are 
computed for the identical set of parameter values and physical systems as described in Fig.~\ref{tran}. Here we fix the decay time 
$\tau=1\,$fs.}
\label{dyna}
\end{figure}
following our analytical discussion, here in Figure~\ref{lhtran} we show the numerical results of the left-handed DSHGs, 
similar to what are presented in Fig.~\ref{tran} for the right-handed ones. For the left-handed DSHG, we need to satisfy two conditions 
associated with $\varphi_{\nu}$ and $\varphi_{\lambda}$, unlike the right-handed one, to have the absolute conducting phase as given in 
Eqs.~\ref{equ26a} and \ref{equ26b}. Because of the sign reversal in the arguments of two cosine modulations (see Eqs.~\ref{equ23} and 
\ref{equ24}), we need to impose the additional restriction to wash out these modulations which yields the absolute conducting phase. 
This is exactly what is reflected from Fig.~\ref{lhtran}(b). When $\varphi_d=\varphi_{\lambda}=0$, we get the 
localized phase for the SRH DSHG (Fig.~\ref{lhtran}(a)), as the transmission probability vanishes completely for all the energy channels. 
In the intermediate phase values the mixture of both conducting and localized phases is obtained (Fig.~\ref{lhtran}(c)). Under this 
condition mobility edge phenomenon is visible, which thus, provides the possibilities of getting LTD transition at selective energies.

For the long-range hopping system, finite transmission is always obtained at some energy channels irrespective of $\varphi_{\nu}$ and 
$\varphi_{\lambda}$ (right column of Fig.~\ref{lhtran}), like the right-handed DSHG. Obviously, for the typical condition where 
$\varphi_d=\varphi_{\lambda}=\pi/2$, absolute conducting phase is generated for the entire energy window (Fig.~\ref{lhtran}(e)). The 
numerical results shown in Fig.~\ref{lhtran} exactly corroborate the analytical description.

\subsection{Role of helical dynamics on electronic localization: Right- and left-handed DSHGs}

Finally, we include the effect of helical dynamics on electronic localization, as static picture does not always give the complete 
\begin{figure}[ht]
{\centering \resizebox*{8.4cm}{8cm}{\includegraphics{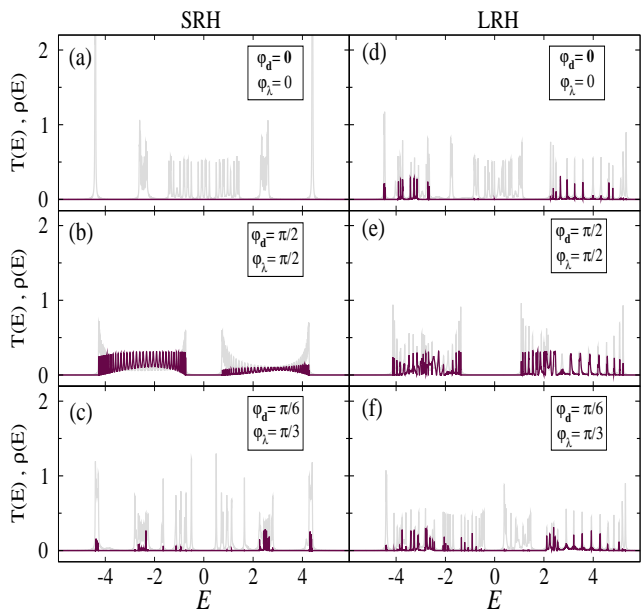}}\par}
\caption{(Color online). Effect of helical dynamics in left-handed DSHG is considered. The transmission probabilities and ADOS are 
evaluated for the identical set of parameter values and physical systems as described in Fig.~\ref{lhtran}. The decay time is 
fixed at $\tau=1\,$fs.}
\label{LHdyna}
\end{figure}
scenario~\cite{hdn1,hdn2,hdn3,hdn4,hdn5,hdn6}. The dynamical effect is incorporated into the system following the methodology introduced 
originally by Ratner and his group~\cite{hdn1}. It has been mentioned that the site energy of the lattice site which is directly coupled 
to the drain electrode gets modified by an imaginary term. It reads as~\cite{hdn1} 
$\epsilon_{j,n}\rightarrow \epsilon_{j,n}-\zeta \hbar/\tau$, where $\tau$ is the decay time and $\zeta=\sqrt{-1}$. The choice of $\tau$ 
is quite important. It should be fixed in such a way that the electron vanishes immediately when it reaches to the drain, but care should 
be taken that $\tau$ cannot be so small that electron gets reflected from the drain end. This methodology for the incoporation of helical 
dynamics is very simple and elegant.

In Figs.~\ref{dyna} and \ref{LHdyna} we present the variations of transmission probabilities and ADOS for the right- and left-handed 
DSHGs, respectively. The results are worked out for the identical systems considering the same parameter values as taken in 
Figs.~\ref{tran} and \ref{lhtran} to make a comparison of localization behaviors between the static and dynamical cases. All the physical 
phenomena i.e., insulating phase, LTD transition, appearance of mixed phase remain unchanged even when the dynamical effect is incorporated. 
Thus, we can say that the localization properties studied here are robust and valid both for the static and dynamic cases.

\section{Closing remarks}

To conclude, in the present article we have theoretically investigated localization properties of double-stranded helical geometries
in presence of transverse electric field. Beacause of this field, site energies get modulated in the form of well known AAH model 
yielding the DSHG as an AAH ladder, even without explicitely considering AAH potentials in the absence of electric field. To mimic
the DSHG with real biological samples like DNA and others, a similar kind of cosine modulation has been introduced in the inter-chain
hopping integrals. Two different cases of hopping, SRH and LRH, have been taken into account, and the interplay between them has been 
critically analyzed. The effects of helical dynamics and the helicity have also been discussed.

Simulating the DSHGs within a tight-binding framework we have investigated localization phenomena under different physical
conditions. For some typical values of the phase factors associated with cosine modulations, analytical results have been worked out,
and later numerical results have been given for more general set of parameter values where analytical solution is no longer possible.
The key aspects and new findings of our work are as follows. 

$\bullet$ The interplay between $\varphi_{\nu}$ and $\varphi_{\lambda}$ is very promising. For the SRH DSHG a complete LTD transition
can be made at some typical values of $\varphi_d$, $\varphi_{\nu}$ and $\varphi_{\lambda}$. At intermediate phase values, mixed states 
appear which yield mobility edge phenomenon. For the other DSHG where long range hoppings are associated, complete localization cannot 
be observed.

$\bullet$ Helicity has an important role on localization. Because of the helicity, site energies are modulated in this cosine form. 
Thus, considering real biological molecules these phenomena can be examined in the presence of external electric field.

$\bullet$ The notable fact is that, for the SRH geometry all the three conducting phases (metallic, insulating and mixed) 
can be achieved in a single system, simply by adjusting $\varphi_{\nu}$ and $\varphi_{\lambda}$. In conventional systems, these three 
conducting phases are usually not observed.

$\bullet$ A comprehesnsive analysis has been given both for the right- and left-handed geometries.

$\bullet$ The localization properties remain unchanged even when the helical dynamics is considered.

At the end, we would like to point out that the characteristic features studied here have not been explored so far to the best of 
our knowledge, and might be interesting to examine the localization phenomena in different chiral biological molecules.

\section*{ACKNOWLEDGMENTS}

SS is thankful to CSIR, India (File number: 09/093(0183)/2017-EMR-I) for providing her research fellowship. The research of SKM is 
supported by DST-SERB, India, under the Project Grant Number EMR/2017/000504. SKM acknowledges fruitful discussions with Prof. S. Sil.
SS and SKM thank all the reviewers for their constructive criticisms, suggestions and valuable comments to improve the quality of the work.

\end{document}